\documentstyle[12pt,epsf]{article}

\begin{document}

\def\CA{{\cal A}}
\def\CB{{\cal B}}
\def\CC{{\cal C}}
\def\CD{{\cal D}}
\def\CE{{\cal E}}
\def\CF{{\cal F}}
\def\CG{{\cal G}}
\def\CH{{\cal H}}
\def\CI{{\cal I}}
\def\CJ{{\cal J}}
\def\CK{{\cal K}}
\def\CL{{\cal L}}
\def\CM{{\cal M}}
\def\CN{{\cal N}}
\def\CO{{\cal O}}
\def\CP{{\cal P}}
\def\CQ{{\cal Q}}
\def\CR{{\cal R}}
\def\CS{{\cal S}}
\def\CT{{\cal T}}
\def\CU{{\cal U}}
\def\CV{{\cal V}}
\def\CW{{\cal W}}
\def\CX{{\cal X}}
\def\CY{{\cal Y}}
\def\CZ{{\cal Z}}

\newcommand{\todo}[1]{{\em \small {#1}}\marginpar{$\Longleftarrow$}}
\newcommand{\ads}[1]{{\rm AdS}_{#1}}
\newcommand{\SL}[0]{{\rm SL}(2,\IR)}
\newcommand{\cosm}[0]{\Lambda}
\newcommand{\tL}[0]{\bar{L}}
\newcommand{\hdim}[0]{\bar{h}}
\newcommand{\bw}[0]{\bar{w}}
\newcommand{\bz}[0]{\bar{z}}
\hyphenation{Min-kow-ski}
\hyphenation{Lor-entz-i-an}
\hyphenation{bound-ary}
\hyphenation{bulk-bound-ary}

\def\pref#1{(\ref{#1})}

\def\ie{{\it i.e.}}
\def\eg{{\it e.g.}}
\def\cf{{\it c.f.}}
\def\etal{{\it et.al.}}
\def\etc{{\it etc.}}

\def\adv{{\it Adv. Phys.}}
\def\ap{{\it Ann. Phys, NY}}
\def\cqg{{\it Class. Quant. Grav.}}
\def\cmp{{\it Commun. Math. Phys.}}
\def\jetp{{\it Sov. Phys. JETP}}
\def\jetpl{{\it JETP Lett.}}
\def\jp{{\it J. Phys.}}
\def\ijmp{{\it Int. J. Mod. Phys. }}
\def\nc{{\it Nuovo Cimento}}
\def\np{{\it Nucl. Phys.}}
\def\mpl{{\it Mod. Phys. Lett.}}
\def\pl{{\it Phys. Lett.}}
\def\pr{{\it Phys. Rev.}}
\def\prl{{\it Phys. Rev. Lett.}}
\def\prcl{{\it Proc. Roy. Soc.} (London)}
\def\rmp{{\it Rev. Mod. Phys.}}
\def\dash{-----------------    }

\def\p{\partial}

\def\apr{\alpha'}
\def\str{{str}}
\def\lstr{\ell_\str}
\def\gstr{g_\str}
\def\Mstr{M_\str}
\def\lpl{\ell_{pl}}
\def\Mpl{M_{pl}}
\def\varep{\varepsilon}
\def\del{\nabla}
\def\tr{\hbox{tr}}
\def\perp{\bot}
\def\half{\frac{1}{2}}
\def\p{\partial}

\def\IB{\relax\hbox{$\inbar\kern-.3em{\rm B}$}}
\def\IC{\relax\hbox{$\inbar\kern-.3em{\rm C}$}}
\def\ID{\relax\hbox{$\inbar\kern-.3em{\rm D}$}}
\def\IE{\relax\hbox{$\inbar\kern-.3em{\rm E}$}}
\def\IF{\relax\hbox{$\inbar\kern-.3em{\rm F}$}}
\def\IG{\relax\hbox{$\inbar\kern-.3em{\rm G}$}}
\def\IGa{\relax\hbox{${\rm I}\kern-.18em\Gamma$}}
\def\IH{\relax{\rm I\kern-.18em H}}
\def\IK{\relax{\rm I\kern-.18em K}}
\def\IL{\relax{\rm I\kern-.18em L}}
\def\IP{\relax{\rm I\kern-.18em P}}
\def\IR{\relax{\rm I\kern-.18em R}}
\def\IZ{\relax{\rm Z\kern-.5em Z}}

\rightline{HUTP-98/A028}
\rightline{CALT68-2171}
\rightline{hep-th/9805171}
\vskip 1cm


\centerline{\Large \bf Bulk vs. Boundary Dynamics} 
\centerline{\Large \bf in Anti-de Sitter Spacetime} 

\vskip 1cm

\renewcommand{\thefootnote}{\fnsymbol{footnote}}
\centerline{{\bf \large Vijay
Balasubramanian${}^{1}$\footnote{vijayb@pauli.harvard.edu},  
Per Kraus${}^{2}$\footnote{perkraus@theory.caltech.edu} 
and Albion Lawrence${}^{1}$\footnote{lawrence@string.harvard.edu}}}
\vskip .5cm
\centerline{${}^1$\it Lyman Laboratory of Physics, Harvard University}
\centerline{\it Cambridge, MA 02138, USA}
\vskip .5cm
\centerline{${}^2$ \it California Institute of Technology}
\centerline{\it Pasadena, CA 91125, USA}

\setcounter{footnote}{0}
\renewcommand{\thefootnote}{\arabic{footnote}}

\begin{abstract}
We investigate the details of the bulk-boundary correspondence in
Lorentzian signature anti-de Sitter space.  Operators in the boundary
theory couple to sources identified with the boundary values of
non-normalizable bulk modes.  Such modes do not fluctuate  and
provide classical backgrounds on which bulk excitations propagate.
Normalizable modes in the bulk arise as a set of saddlepoints of the
action for a fixed boundary condition.  They fluctuate and describe
the Hilbert space of physical states.  We provide an explicit,
complete set of both types of modes for free scalar fields in global
and Poincar\'e coordinates.  For $\ads{3}$, the normalizable and
non-normalizable modes originate in the possible representations of
the isometry group $\SL_L\times\SL_R$ for a field of given mass.  We
discuss the group properties of mode solutions in both global and
Poincar\'e coordinates and their relation to different expansions of
operators on the cylinder and on the plane.  Finally, we discuss the
extent to which the boundary theory is a useful description of the
bulk spacetime.
\end{abstract}

\def\figloc#1#2{\bigskip\vbox{{\epsfxsize=4.5in \nopagebreak[3]
        \centerline{\epsfbox{diag#1.ps}} \nopagebreak[3] \centerline{Figure
        #1} \nopagebreak[3] {\raggedright\it \vbox{ #2 }}}} \bigskip }

\section{Introduction}

The description of certain charged black holes as D-branes in string
theory implies a connection between the low-energy gauge dynamics on
the brane and the low energy supergravity in
spacetime.~\footnote{Amongst many other works,
see~\cite{oldgaugerefs}.}  Recently, Maldacena has proposed decoupling
limits in which the brane gauge dynamics is dual to string theory on
the near-horizon anti-de Sitter geometry of the corresponding black
hole~\cite{juanads}.

A more precise definition of this duality was developed in~\cite{gkp,
holowit}.  We associate to the string compactification on
$\ads{d+1}\times\CM$ a conformal field theory living on a space
conformal to the $d$-dimensional boundary $\CB$ of the AdS factor.  To
each field $\Phi_i$ there is a corresponding local operator $\CO^i$ in
the conformal field theory.  The relation between string theory in the
bulk and field theory on the boundary is:
\begin{equation}
	Z_{{\rm eff}} (\Phi_{i}) = e^{i S_{{\rm eff}}(\Phi_{i})}
		= \langle T e^{i \int_\CB \Phi_{b,i} \CO^i } \rangle\ .
\label{eq:bbrel}
\end{equation}
Here $S_{{\rm eff}}$ is the effective action in the bulk, $\Phi_{b,i}$
is the field $\Phi_i$ restricted to the boundary, and $T$ is the
time-ordering symbol in the field theory on $\CB$.  The expectation
value on the right hand side is taken in the boundary field theory,
with $\Phi_{b,i}$ treated as a source term.  In the classical
supergravity limit, given a boundary field we solve for the
corresponding bulk field and use it to relate the bulk effective
action to boundary correlation functions.\footnote{By ``boundary
correlation functions'' we mean the correlation functions of the CFT.}

In Euclidean AdS this proposal used the absence of normalizable
solutions to the field equations, and the resulting unique extension
of a boundary field $\Phi_{b,i}$ into the bulk~\cite{holowit}. We are
interested in issues of spacetime causal structure and dynamics, so we
would also like a Hamiltonian formulation of the bulk theory in
Lorentzian signature AdS spaces. The standard construction of quantum
field theory depends on the existence of a complete set of
normalizable modes, which is in tension with the unique extendibility
of AdS boundary conditions into the bulk.  Indeed, several consistent
quantizations have been found~\cite{avisetal, brfreed}, involving
particular choices of boundary conditions for AdS spacetimes and the
resulting set of normalizable modes.  On the other hand, the
bulk-boundary correspondence demands the ability to tune the boundary
conditions in order to describe the appropriate boundary correlation
functions.  In this paper we resolve these tensions by arguing that
the bulk-boundary correspondence as formulated
in~\cite{juanads,gkp,holowit} demands the inclusion of both
normalizable and non-normalizable modes.  The former propagate in the
bulk and correspond to physical states while the latter serve as
classical, non-fluctuating backgrounds and encode the choice of
operator insertions in the boundary theory.

We begin in Sec.~\ref{sec:review} by reviewing the computation of
boundary correlation functions, and providing a prescription for
computing the bulk effective action in a Hamiltonian formulation.  We
will argue that specifying the boundary conditions involves turning on
non-normalizable modes which do not fluctuate.  Including such
non-fluctuating modes may seem strange, but several well-known
examples exist in other contexts: a classic example is a field theory
which undergoes spontaneous symmetry breaking, and more recent related
examples arise in \cite{callanwilczek, notesliouv, backind}.
Normalizable solutions to the wave equation are then used in the mode
expansion of operators and the construction of a Fock space.  The
resulting Hilbert space of states is identified with the Hilbert space
of the boundary theory~\cite{garyhirosi,holowit}.
In Sec.~\ref{sec:norm} we make this more explicit by studying
solutions to the wave equation for free scalars of arbitrary mass.  We
find that for general masses, the field equations have both
normalizable and non-normalizable solutions and the latter couple to
the boundary.  

In Sec.~\ref{sec:ads3} we specialize to $\ads{3}$, whose bulk isometry
group and boundary conformal group are both $G = \SL_L\times\SL_R$.
We show that the normalizable modes transform in
unitary irreducible representations of $G$ while the non-normalizable modes
transform in non-unitary reducible representations which contain a highest weight
module.  The normalizable and non-normalizable highest weight
representations are built on states with $\SL_L\times\SL_R$ weights
$h_L = h_R = h_{\pm} = (1/2)(1 \pm \sqrt{m^2\cosm^2 +
1})$.\footnote{Here $\cosm$ is the inverse of the cosmological
constant of $\ads{3}$.}  Interesting subtleties arise for small masses
and for integral $\nu = (1/2) \sqrt{m^2 \cosm^2 + 1}$.  We carry out
the analysis in both global and Poincar\'e coordinates in order to
discuss conformal field theories both on the cylinder and the plane.
The latter case has some curious features because Poincar\'e
coordinates only cover a patch of the global spacetime.  We conclude
the paper with a discussion of the utility and limitations of the
bulk-boundary correspondence for describing bulk spacetime physics via
the boundary gauge theory.

\section{Boundary correlators from the bulk}
\label{sec:review}

\paragraph{Euclidean formulation: }  Specifying the
boundary behaviour of a field $\Phi$ in Euclidean AdS leads to a
unique solution to the equations of motion, given some regularity
conditions (see~\cite{holowit} for discussion and references).  So
(\ref{eq:bbrel}) is unambiguously interpreted by evaluating the
$\ads{d+1}$ effective action on the unique bulk extension of the
boundary field.  For a free scalar $\Phi$ with boundary value
$\Phi_b$,
\begin{equation}
	S_{{\rm eff}} = \int _{{\rm AdS}_{d+1}} d^{d+1} x \sqrt{g}
\left(g^{\mu\nu} \p_\mu \Phi \p_\nu \Phi + m^2 \Phi^2 \right)\ .
\end{equation}
We relate this to the right hand side of (\ref{eq:bbrel}) by integrating
by parts.  Since $\Phi$ is a solution to the equations of motion, the bulk
contribution vanishes.  There is, however, a boundary term which is
non-vanishing for the relevant solutions to the Euclidean wave equation:
\begin{equation}
	S_{{\rm eff}} = \lim_{r \rightarrow 0} \int d\tau d^{d-1} \vec{x}
		\sqrt{g} g^{rr} \Phi \partial_r \Phi\ .
\label{eq:boundpart}
\end{equation}
(Here $r=0$ is the boundary of spacetime in the Poincar\'e coordinates
defined in the Appendix.)
As $r \rightarrow 0$ this is quadratic in $\Phi_b$ and can be thought of
as a quantity in the boundary theory.
In order to compare to the right-hand side of (\ref{eq:bbrel}), we
must identify the boundary operator coupling to $\Phi$.  The dimension of
the operator is determined by the growth of $\Phi$ at the boundary of 
AdS~\cite{gkp, holowit}.  In particular, 
suppose that the boundary behavior of $\Phi$ is
\begin{equation}
	\Phi \buildrel r\rightarrow 0 \over \longrightarrow 
		r^{-\lambda} \Phi_0 (\vec{x})\ ,
\label{eq:boundlim}
\end{equation}
Then the corresponding operator has mass dimension $d+\lambda $.  In
fact, this procedure is not completely well-defined: in general,
(\ref{eq:boundpart}) blows up and some regularization is required in
order to extract the correlator \cite{freedcorr}. For interacting
theories, we can calculate $S_{{\rm eff}}$ perturbatively by
integrating over fluctuations away from the classical solution.  These
fluctuations should have finite action and vanish appropriately at the
boundary.

\paragraph{Lorentzian formulation: } In Lorentzian signature
AdS spacetime the relation in (\ref{eq:bbrel}) is more subtle, because
there are normalizable solutions to the wave equation~\cite{avisetal,
brfreed,fronsdal} which do not affect the leading boundary behavior
(\ref{eq:boundlim}).  So the boundary value $\Phi_b$ does not uniquely
specify the bulk field.  Furthermore, the normalizable modes cannot be
projected out since they are needed to expand quantum operators, build
a Fock space, and compute Green's functions in the bulk.

The bulk effective action
appearing in (\ref{eq:bbrel}) can be written as:
\begin{equation}
Z_{{\rm eff}}(\Phi_i) = \int \CD\Phi_i \, e^{iS(\Phi)}
\label{eq:path1}
\end{equation}
where the path integral\footnote{Here we are restricting attention to
those low energy processes for which a field theory path integral
makes sense.  More generally, the full string theory partition
function is implied.}  is taken over fields $\Phi_i$ that take the
boundary value $\Phi_{b,i}$.  The measure $\CD\Phi_i$ can be
normalized by requiring that $Z_{{\rm eff}} = 1$ when the boundary
condition is trivial ($\Phi_{b,i} = 0$).  In general there is a set of
saddle points that must be summed over; these correspond to the
normalizable modes that are solutions to the equations of motion with
a fixed boundary condition.  For a free theory in Minkowski space, the
normalizable modes indicate that there is a manifold of flat
directions that must be integrated over for fixed boundary behavior.
For a free theory the calculations in Refs.~\cite{gkp,holowit}\ are
not affected as the flat directions lead to an overall volume factor
that is divided out when we appropriately normalize the path integral.

In order to define the path integral in (\ref{eq:path1}) we must
specify the behaviour of fields at the AdS boundary.  We will see that
boundary conditions can be specified by including certain modes that
are generically not normalizable or at least perturb the asymptotic
geometry too violently.  In fact, these are precisely the modes that
are identified in \cite{gkp,holowit}\ with sources coupling to
boundary operators.  As such, these modes should be locked or
non-fluctuating.  Such non-fluctuating modes arise in a variety of
contexts.  One example is the homogeneous mode of a scalar field that
takes a nonzero value through spontaneous symmetry breaking.  Another
example is Euclidean field theory on spaces with constant negative
curvature; the authors of ref. \cite{callanwilczek} note that the
large volume at infinity keeps modes with certain boundary conditions
from fluctuating.  Our discussion is particularly reminiscent of Liouville
theory, where local operators create non-normalizable wavefunctions
when inserted in a disc amplitude \cite{notesliouv,backind}.  This
analogy is especially attractive after the work in \cite{emilmatrix}\
and was a motivation for the present form of this paper; indeed, for
$\ads{3}$ the gravity sector of the bulk has a Chern-Simons action
which reduces to a boundary Liouville theory~\cite{vandriel, banados}.
In particular ref. \cite{emilmatrix}\ relates the asymptotic behavior
of solutions to the wave equation to the gravitational dressing of
chiral operators in the boundary CFT.

For an interacting theory, our prescription provides the analog of an
S-matrix for field theories in AdS space, an issue raised in
Ref.~\cite{banksgreen}.  In standard flat-space field theory, one
specifies boundary conditions at infinity by turning off the
interactions and physically separating the initial excitations, so
that we can sensibly discuss asymptotic states.  In our picture, even
though the bulk excitations may not be separable and asymptotic states
may not exist, we can sensibly discuss the dependence of the partition
function on boundary conditons for locked fields at infinity.

\paragraph{Hamiltonian quantization: }
Once we have specified the boundary conditions, we must ask what
fluctuating modes to keep in a Hamiltonian formalism.  The point is
that quantization requires a complete, normalizable set of field
modes, which in AdS space requires some sort of boundary condition at
infinity.  Several consistent choices have been suggested in the past;
for the bulk-boundary correspondence to make sense the choice should
be unambiguously determined by the physics of the situation.  Our
prescription is heavily motivated by the work of Breitenlohner and
Freedman \cite{brfreed} (which covers $\ads{4}$), so we will review
their discussion for scalars of arbitrary mass $m$ in $\ads{d+1}$
(also see \cite{meztown}).

Fix a spacelike slice $\Sigma \subset {\rm AdS}_{d+1}$ with
coordinates $x$ and an orthogonal, timelike coordinate $t$.  Given two
solutions $u_1,u_2$ to the scalar wave equation, define the inner
product:
\begin{equation}
	(u_1, u_2) = i\int _\Sigma d^d x \sqrt{g} g^{t t}
		\left(u_1^{\ast}\p_t u_2 - \p_t u_1^{\ast} \, u_2 \right)\ .
\label{eq:adsnorm}
\end{equation}  
If $u_1=u_2$, this is the integral of the time component of the
current
\begin{equation}
	j^{\mu} = i g^{\mu\nu} \left( u^{\ast} \p_\nu u -
		\p_\nu u^{\ast} u \right)\ .
\end{equation}
Here $j^t$ is only time-independent up to boundary terms coming from
the timelike boundary of anti-de Sitter space; it is easy to see,
using the equations of motion, that
\begin{equation}
	\p_t j^t = \int_{\p\Sigma} dA_k \, \, j^k\ ,
\label{eq:boundpr}
\end{equation}
where $\p\Sigma$ is the boundary of the spacelike slice.  The authors
of~\cite{brfreed}\ find complete sets of modes by demanding, first,
that the flux normal to the boundary vanish at infinity.  They also
demand that no energy is exchanged with the boundary:
\begin{equation}
	F = \int_{\p\Sigma} dA_k \sqrt{g} \, T^k_0 = 0\ .
\end{equation}
The authors of~\cite{brfreed,meztown} considered the relative merits
of the canonical and improved stress tensors in this equation.  This
potential ambiguity in the stress tensor arises because the curvature
is constant, and its possible coupling to a scalar field would appear
as an effective mass term. In our case the choice is dictated by the
underlying string theory and the supergravity effective action that
descends from it.  We will simply treat scalar fields of mass squared
$m^2$ with the understanding that the mass term may contain a
contribution from a curvature coupling.

We will find that for masses $m^2 \geq 1 - \frac{d^2}{4}$, simply
requiring normalizability is sufficient to isolate a suitable class of
fluctuating solutions, and the more refined discussion in
Ref.~\cite{brfreed}\ is unnecessary.  For $-d^2/4 < m^2 < 1 - d^2/4$
(stability requires $m^2 \geq -d^2/4$), there are two sets of
normalizable solutions and some criterion is needed to distinguish
them.  Refs. \cite{brfreed,meztown} show that for any given field
propagating on $\ads{d\geq4}$, conservation of the inner product
(\ref{eq:adsnorm}) and the vanishing of $F$ at the boundary requires
either but not both sets.  Furthermore, in the case of $\ads{4}$, the
authors of \cite{brfreed} show that the two sets of modes are built on
representations of the conformal algebra with different lowest energy
states and that supersymmetry generically requires one to take both
types of mode into account.  The clearest example is that of the
scalar and pseudoscalar in the gravity supermultiplet of gauged $N=4$
supergravity \cite{brfreed}.  Both are conformally coupled, and
supersymmetry requires that the modes of the scalar lie in one
representation while the modes of the pseudoscalar lie in the other.
Further criteria are needed, however, to decide which assignment is
realized.  In this particular case, Hawking \cite{hawkads}\footnote{We
would like to thank S. Ross for pointing out this reference and its
relevance.}  imposed the requirement that the
metric be asymptotically anti-de Sitter (in a sense defined in that work)
to find particular boundary conditions on the linearized metric
perturbations.  Supersymmetry then requires the scalar mode to
live in the representation whose highest-weight state has the lowest energy.
It would be nice if similar criteria could be applied to general modes
in the range $-d^2/4 < m^2 < 1 - d^2/4$, independently of supersymmetry;
we will not investigate this point, however.

\paragraph{Summary: }The lesson is that we need to keep both normalizable
and non-normalizable modes in AdS spacetimes. The non-normalizable modes
correspond to operator insertions in the boundary gauge theory;
from the AdS point of view they provide non-trivial boundary conditions. 
The normalizable modes fluctuate in the
bulk; quanta occupying such modes have a dual description in
the boundary Hilbert space.   In path integral language,
(\ref{eq:bbrel}) 
can be understood via the background field expansion: we compute the
effective action by expanding the path integral of the bulk theory as
\begin{equation}
	\Phi = \Phi_{cl} + \delta\Phi\ .
\end{equation}
$\Phi_{cl}$ is a classical, nonnormalizable solution to the equations of
motion, corresponding to an operator insertion at infinity and a particular
choice of boundary conditions.  Then $\delta\Phi$ is the fluctuating piece
over which we integrate to get the partition function.  
The normalizable modes appear as stationary points of the action given the
background $\Phi_{{\rm cl}}$.

\section{Explicit modes for scalars of arbitrary mass}
\label{sec:norm}
In this section we show that a scalar of arbitary mass in AdS
spacetime has both fluctuating and non-fluctuating solutions that
implement the bulk-boundary correspondence as advocated above.   
Generically, the non-fluctuating solutions are not normalizable in 
the norm (\ref{eq:adsnorm}).   

\subsection{Solutions in Poincar\'e coordinates}
\label{sec:poinsoln}
We begin with solutions in Poincar\'e coordinates,
which allows for a direct comparison with refs.~\cite{gkp, holowit}.
It is easy to separate variables
in these coordinates by writing the scalar of mass squared 
$m^2/\Lambda^2$ as:
\begin{equation}
	\Phi = e^{-i\omega\tau + i\vec{k}\cdot\vec{x}}
		r^{\frac{d}{2}} \chi(r)\ .
\end{equation}
$\chi$ then satisfies the equation
\begin{equation}
    r^2 \,\p_r^2 \chi + r\,\p_r \chi - \left((m^2 + \frac{d^2}{4}) + (
		\vec{k}^2 - \omega^2)\,r^2 \right) \chi \ = 0.
\end{equation}

For $q^2 = \vec{k}^2 - \omega^2 > 0$, the solution is ~\cite{gkp, freedcorr}:\footnote{
We use the notation in \cite{absteg} for Bessel functions.}
\begin{equation}
	\Phi^{{\rm s.l.}} = e^{-i\omega t + i \vec{k}\cdot \vec{x}}
		r^{\frac{d}{2}} K_{\nu}(q r)\ ,
\end{equation}
where $\nu=\half\sqrt{d^2 + 4m^2}$.
This solution is non-normalizable at the boundary
at infinity but well-behaved in the interior.   The second,
independent solution $I_\nu$ is very badly behaved (i.e. blows up
exponentially) in the interior and is therefore eliminated.
If we consider anti-de Sitter
space as the near-horizon geometry of a brane, then in
the asymptotically flat region $\Phi^{{\rm s.l.}}$ would have imaginary
momentum perpendicular to the brane.

For $q^2 < 0$, there are two possible solutions which are 
regular in the interior.  If $\nu$ is not integral,
\begin{equation}
	\Phi^{(\pm)} = e^{-i\omega t + i \vec{k}\cdot \vec{x}}
		r^{\frac{d}{2}} J_{\pm\nu}(|q| r)
\label{eq:poinnorm}
\end{equation}
are two independent solutions.  If $\nu$ is integral, $J_{\pm \nu}$ are
equivalent, and the two independent solutions
are $\Phi^{(+)}$ in (\ref{eq:poinnorm}) and
\begin{equation}
	\Phi^{(-)} = e^{-i\omega t + i\vec{k}\cdot\vec{x}}
		r^{\frac{d}{2}} Y_\nu (|q| r)\ .
\end{equation}
(Note that many of the scalar modes arising from the Kaluza-Klein
reduction of string theory down to ${\rm AdS_{d+1}}$ have integral
$\nu$
\cite{holowit,biranetal,casteletal,vann,romansetal,sezginetal,oferetal},
so this is a ``special case'' of particular importance.)  The boundary
term (\ref{eq:boundpart}) will vanish for $\Phi^{(+)}$; for
$\Phi^{(-)}$ it will either go to a constant (for $m^2 = 0,d=2$) or
blow up.  $\Phi^{{\rm s.l.}}$ and $\Phi^{(-)}$ both behave as
\begin{equation}
	\Phi \sim r^{2h_-} \Phi_0
\end{equation}
as discussed in \cite{gkp, holowit}, where 
\begin{equation}
	h_\pm = \frac{d}{4} \pm \frac{1}{4} \sqrt{d^2 + 4m^2}\ .
\end{equation}
For $\nu = 0$, $\Phi^{(-)}$ behaves as
\begin{equation}
	\phi \sim r^{\frac{d}{2}}\ln r\ .
\end{equation}
As in \cite{gkp, holowit}, $\Phi^{(-,{\rm s.l.})}$ couple to operators of
dimension $2h_+$.

The asymptotic behavior of Bessel functions is well-known, so criteria
for selecting fluctuating and non-fluctuating modes are easy to apply.
For $\nu > 1$, only $\Phi^{(+)}$ is normalizable and $\Phi^{(-)}$ must
therefore act as the non-fluctuating source term in (\ref{eq:bbrel}).
For $\nu < 1$ the story is more complicated as both of $\Phi^{(\pm)}$
are normalizable and, as pointed out in Refs. \cite{brfreed,meztown},
both kinds of modes are necessary for supersymmetry.  Indeed, for
compactifications of M-theory on $\ads{4}\times S^7$, a conformally
coupled scalar and pseudoscalar appear in the KK spectrum
\cite{biranetal,casteletal,oferetal,minads}.\footnote{We would like to
thank O. Aharony for pointing out this example, and for discussing the
$\nu < 1$ case in general.}  Both analysis of the KK spectrum and its
relation to operators in the dual 3D CFT \cite{oferetal,minads}, and
the arguments in \cite{hawkads}, show that the fluctuating modes for
the scalar are of the form $\Phi^{(-)}$ and the fluctuating modes for
the pseudoscalar are of the form $\Phi^{(+)}$.  Furthermore, the
``conjugate'' solutions $\Phi^{(+)}$ and $\Phi^{(-)}$
couple to operators with dimension $h_-$ and $h_+$, respectively
\cite{oferetal,minads}.\footnote{Note that the fact that the scalar
mode has dimension $1 < \frac{d}{2}$ contradicts the statement made 
in Ref. \cite{holowit} for Euclidean theories
that the dimension of operators in the CFT is
bounded below by $\frac{d}{2}$.  It however satisfies the unitarity
condition $h \geq \half(d-2)$ for scalar operators in the dual CFT
\cite{mincft}.}  One may worry legitimately that for the scalar mode,
interpreting (\ref{eq:bbrel}) is problematic because the surface term
(\ref{eq:boundpart}) diverges for this mode and not for the ``source''
modes coupling to the boundary.  Perhaps the answer is that when
computing correlation functions, one does not perturb the background
with classical normalizable modes.  At present we will leave the
resolution of this issue for future work.

It is also worth noting that the borderline non-normalizable modes,
$\Phi^{(-)}$ for $h_- = \half(d-2)$, contain the singleton
\cite{singleton,ferrara-one}.

\subsection{Solutions in global coordinates}
\label{sec:globsoln}
Let us examine a scalar of mass $m^2/\Lambda^2$.
Make the substitution:
\begin{equation}
\label{eq:globalsep}
	\Phi = e^{-i\omega t} Y_{\ell,\{m\}} (\Omega) \chi(\rho)\ .
\end{equation}
Here $Y_{\ell}$ are the $\ell^{\rm th}$ spherical harmonics on $S^{d-1}$,
for which:
\begin{equation}
	\nabla^2_{S^{d-1}} Y_{\ell} = -\ell (\ell + d - 2) Y_{\ell}\ .
\end{equation}
with $\ell \geq 0$.  The wave equation in global coordinates (see the
Appendix) is:
\begin{equation}
\label{eq:globalwave}
	\frac{1}{(\tan\rho)^{d-1}} \p_{\rho} \left( (\tan\rho)^{d-1}
                                               \p_{\rho} \right)
		\chi + \left[ \omega^2 - \ell(\ell + d - 2)\csc^2\rho
		- m^2 \sec^2 \rho \right] \chi = 0\ .
\end{equation}

We have chosen $\Phi$ with which we can build an arbitrary
configuration on $S^{d-1}\times \IR$ for a given $\rho$;
(\ref{eq:globalwave})\ then gives us the $\rho$ dependence.  For
general $m^2$, $\omega$ the equation of motion is easily converted
into a hypergeometric equation which can naturally be expanded in
variables that vanish either at the origin ($\rho=0$) or the boundary
($\rho=\pi/2$).  For maintaining regularity at the origin, it will be
easier to examine the solutions as functions of $\sin^2 \rho$.  For
examining the boundary behavior, solutions as functions of
$\cos^2\rho$ will be more convenient.  We will display both sets of
solutions explicitly.  Of course one can pass between them using
standard formulae relating hypergeometric functions as arguments of
$z$ and $1-z$~\cite{absteg, erdelyi, gandr}; these permit the
enforcement of regularity conditions at the origin and the boundary ,
thereby imposing quantization conditions on the spectrum of the
theory.

Begin by substituting:
\begin{equation}
	\chi(\rho) = (\cos\rho)^{2h} (\sin\rho)^{2b} f(\rho)\ .
\end{equation}
Let $y = \sin^2\rho$.  It is then easy to see that:
\begin{equation}
	y(1-y)\partial_y^2 f + \left[2b + \frac{d}{2} - (2h + 2b + 1) y\right]
		\partial_y f - \left[ (h + b)^2 - \frac{\omega^2}{4} \right] f = 0\ ,
\label{eq:hypersinewave}
\end{equation}
where
\begin{equation}
	h(h - \frac{d}{2}) = \frac{m^2}{4};\ \ \ 
2b(2b + d - 2) = \ell(\ell + d - 2)\ .
\end{equation}
The equations for $h,b$ each have two solutions:\footnote{
Let $\lambda^{{\rm bf}}$ be what ref.~\cite{brfreed}\ calls $\lambda$
and $\lambda^w_\pm$ be what ref.~\cite{holowit}\ calls $\lambda$.
Then $h$ is related to these as 
$h_\pm = \half \lambda^{{\rm bf}}_\pm = - \half \lambda^w_\mp$}
\begin{eqnarray}
	h_\pm &=& \frac{d \pm \sqrt{d^2 + 4m^2}}{4}\\
	b &=& \frac{\ell}{2}\ ,\ \frac{1}{2}\left(2 - d - \ell\right)\ .
\end{eqnarray}
The hypergeometric equation will have two independent solutions,
corresponding to the two solutions of the indicial equation for $b$.
One solution will be logarithmic if $\ell + \frac{d}{2}$ is an
integer, \ie\ if $d$ is even.

Choosing instead $x = \cos^2\rho$, one gets the hypergeometric
equation:
\begin{equation}
	x(1-x) \p_x^2 f + \left[ 2h + 1 - \frac{d}{2} -
		\left(2h + 2b + 1\right) x \right] \p_x f
		+ \left[ \left(h + b\right)^2 - \frac{\omega^2}{4} \right] f = 0\ ,
\label{eq:hyperwave}
\end{equation}
with $h,b$ as before.  Again, the hypergeometric equation has two
independent solutions; this time they correspond to the two solutions
of the indicial equation for $h$.  One solution will be logarithmic if
\hbox{$\nu=\frac{1}{2}\sqrt{d^2+4m^2}$} is an integer.  In fact, if we
wish to transform solutions as functions of $\sin^2\rho$ to solutions
as functions of $\cos^2\rho$, the relevant formulae are modified in
the case of integral $\nu$; thus we will examine the two cases
separately.

\paragraph{Behaviour at the origin: } The behaviour at the origin 
is conveniently analyzed by studying solutions as a function of
$\sin^2\rho$.   Choose, without loss of generality, $h=h_+$.  The
first solution as a function of $\sin^2\rho$ is:\footnote{We use the
notation in \cite{absteg}.}
\begin{eqnarray}
	\lefteqn{\Psi^{(1)} = e^{-i\omega t}Y_{\ell}(\Omega)
		\, (\cos\rho)^{2h_+}\, (\sin\rho)^{\ell}}\nonumber\\
	&&_2F_1(h_{+} + \half(\ell + \omega),
		h_+ + \half(\ell - \omega), \ell + \frac{d}{2}; \sin^2\rho)\ .
\label{eq:psione}
\end{eqnarray}
The second solution depends on whether $d$ is even or odd.
If $d$ is odd, then:
\begin{eqnarray}
	\lefteqn{\Psi^{(2)} = e^{-i\omega t}Y_{\ell}(\Omega)
		\, (\cos\rho)^{2h_+}\, (\sin\rho)^{2-d-\ell}}\nonumber\\
	&&_2F_1(h_+ +\half(2 - d - \ell +\omega),
		h_+ + \half(2 - d - \ell - \omega),
		2 - \ell - \frac{d}{2}; \sin^2\rho)\ .
\label{eq:psitwo}
\end{eqnarray}
If $d$ is even the second solution is logarithmic:
\begin{eqnarray}
	\lefteqn{\Psi^{(2)} = \Psi^{(1)} \ln \sin^2\rho +}\nonumber\\
	&&\sum_{k=1}^{\infty} \sin^{2k}\rho \frac{(h_+ + \half(\ell+\omega))_k
		(h_+ + \half(\ell - \omega))_k}{(\ell + \frac{d}{2})_k \,
	k!} 
\times \nonumber \\
           &&
		\left[h(0,\ell+{d\over2} - 1,\nu+1,\omega,k) - 
                h(0,\ell+{d\over2} - 1,\nu+1,\omega,0)\right]
		\nonumber \\
	&&- \sum_{k=1}^{\ell + \half d - 1}
		\frac{(k-1)! (1 - \ell - \frac{d}{2})_k}{(1 - h_+ - 
			\half(\ell + \omega))_k (1 - h_+ - \half(\ell + \omega))_k}
		\sin^{-2k}\rho\ ,
\end{eqnarray}
where
\begin{eqnarray}
	\lefteqn{h(e,f,g,\omega,k) = \psi\left(\half[\half e + f + g +\omega] + k
			\right)} \nonumber \\
		&&+ \psi\left(\half[\half e+ f+ g-\omega] + k\right)
		- \psi(1+ f +k) - \psi(1+k)\ ,
\label{eq:hdef}
\end{eqnarray}
following ref. \cite{gandr}, and
\begin{equation}
	(a)_k = \frac{\Gamma(a+k)}{\Gamma(a)}\, ~~~~~~~~~~
	\psi(x) = \frac{d}{dx} \ln \Gamma(x)\ .
\end{equation}

We must impose a regularity condition at the origin because in order for
(\ref{eq:bbrel}) to make sense, we should not have contributions to
correlation functions coming from the interior.  So we will only keep 
solutions for which the boundary term of the classical action
vanishes at the origin $\rho=0$:
\begin{equation}
	S_{{\rm origin}} = \lim_{\rho \to 0} \int_{\rho\ {\rm fixed}} dtd\Omega
		\sqrt{g}g^{\rho\rho}\Phi\partial_\rho\Phi \to 0\ .
\label{eq:originpart}
\end{equation}
It is easy to show that this means that only the first solution
$\Psi^{(1)}$ is allowed.

\subsubsection{$\nu$ nonintegral}

\paragraph{Behaviour at the boundary: } 
To study the behaviour at the boundary it is most convenient to work with
solutions as a function of $\cos^2\rho$:
\begin{eqnarray}
	\lefteqn{\Phi^{(+)} = e^{-i\omega t} \, Y_{\ell,\{m\}}
		(\Omega) \, (\cos\rho)^{2h_+} \,
		(\sin\rho)^{\ell}}\nonumber\\ &&_2F_1( h_+ + \half
		(\ell + \omega), h_+ + \half(\ell - \omega), 2h_+ + 1
		- \frac{d}{2}; \cos^2\rho )\ ,
\label{eq:phiplus}
\end{eqnarray}
and
\begin{eqnarray}
	\lefteqn{\Phi^{(-)} = e^{-i\omega t} \, Y_{\ell,\{m\}} (\Omega)
		\, (\cos\rho)^{2h_-} \, (\sin\rho)^\ell}\nonumber\\
	&&{}_2F_1(h_-
			+ \half(\ell + \omega), h_- + \half(\ell - \omega),
			2h_- + 1 - \frac{d}{2}; \cos^2\rho)\ .
\label{eq:phiminus}
\end{eqnarray}
In general the regular solution at the origin ($\Psi^{(1)}$) is a linear 
combination of $\Phi^{(\pm)}$.

We can see directly that the leading behavior at the boundary of
$\Phi^{(\pm)}$ is $(\cos\rho)^{2h_\pm}$.  For $\nu > 1$ the norm
(\ref{eq:adsnorm}) of $\Phi^{(-)}$ diverges at $\rho = \frac{\pi}{2}$,
while the norm of $\Phi^{(+)}$ converges.  Thus, up to regularity at the
origin, $\Phi^{(+)}$ is our candidate normalizable mode.  Similarly, a
combination of $\Phi^{(+)}$ and $\Phi^{(-)}$ is a candidate
non-normalizable mode.  Again, its behavior at infinity indicates that it
will couple to operators of dimension $h_+$ in (\ref{eq:bbrel}).

For $\nu < 1$ both modes are well-behaved at infinity
and further examination is required to select
the relevant fluctuating modes,
just as in our discussion of the solutions
in Poincar\'e coordinates.  Again,
fields which have the solutions $\Phi^{(\pm)}$ as
their fluctuating modes will have solutions
$\Phi^{(\mp)}$ which act as source terms
in (\ref{eq:bbrel}) for the related operator of dimension
$h_\pm$ in the dual CFT.

\paragraph{Quantization condition for normalizable modes: }
As we have discussed, regularity at the origin requires the choice 
of $\Psi^{(1)}$.   This solution can be written as a linear combination
of $\Phi^{(\pm)}$:
\begin{equation}
	\Psi^{(1)} = C^{(+)} \Phi^{(+)} + C^{(-)} \Phi^{(-)}\ ,
\end{equation}
where
\begin{eqnarray}
	C^{(+)} &=& \frac{\Gamma(\ell + \frac{d}{2})\Gamma(-\nu)}
		{\Gamma(h_{-} + \half(\ell + \omega))\Gamma(h_{-}
			+ \half(\ell-\omega)}\nonumber\\
	C^{(-)} &=& \frac{\Gamma(\ell + \frac{d}{2})\Gamma(\nu)}
		{\Gamma(h_{+} + \half(\ell + \omega))\Gamma(h_{+}
			+ \half(\ell-\omega)}\ .
\label{eq:lincomb}
\end{eqnarray}
(Note that $\nu$ is assumed to be non-integral here.)
For $\nu > 1$, $C^{(-)}$ must vanish for a fluctuating solution because 
the norm of $\Phi^{(-)}$ diverges at the boundary. 
This will happen if one of the gamma functions in the denominator
has zero or a negative integer as its argument, \ie\ if
\begin{equation}
	\omega = \pm (2h_+ + \ell + 2n); \ \ \ \ \ n = 0,1,2,\ldots\ .
\label{eq:hplusquant}
\end{equation}
So this is the spectrum of normalizable modes $\Phi^{(+)}$ for $\nu > 1$.
For the same range of $\nu$, non-fluctuating modes do not have a
quantization condition.   For a special set of frequencies
\begin{equation}
	\omega = \pm (2h_- + \ell + 2n); \ \ \ \ \ n = 0,1,2,\ldots\ .
\label{eq:nonnormquant1}
\end{equation}
the non-fluctating modes are purely of the $\Phi^{(-)}$ type.  We will see
in the next section that such modes are in a highest weight representation.
In general, however, the non-normalizable modes are simply the linear
combination appearing in (\ref{eq:lincomb}).\footnote{In a previous
preprint version of this work, we imposed a quantization condition on the
non-normalizable solutions also, by demanding that the be purely of the
$\Phi^{(+)}$ type.  In fact, there is no need to impose such a condition.
Since the AdS/CFT correspondence should work when time is noncompact, we
should be allowed sources with arbitrary time dependence.  We would like to
thank J. Maldacena for emphasizing this to us.}

For $\nu < 1$, both $\Phi^{(+)}$ and $\Phi^{(-)}$ are potentially
normalizable because both norms are well-behaved at the boundary.
The quantization condition (\ref{eq:hplusquant}) gives the spectrum of
normalizable modes of the form $\Phi^{(+)}$.
Normalizable modes of the form  $\Phi^{(-)}$ are obtained by  imposing instead:
 \begin{equation}
	\omega = \pm (2h_- + \ell + 2n); \ \ \ \ \ n = 0,1,2,\ldots\ .
\label{eq:hminusquant}
\end{equation}
After picking one of these towers as the fluctuating modes, the remaining
modes that are regular at the origin should be locked and mediate the
bulk-boundary correspondence.  As we have already discussed, the case when
$\nu < 1$ and $\Phi^{(-)}$ is the fluctuating mode is confusing since the
locked mode falls off faster at the boundary than the fluctuating mode.  
As a result, for the non-fluctuating mode to couple as source
to an operator in the boundary theory with dimension $h_-$,
it must not contain any terms which behave as $(\cos\rho)^{2h_-}$
at the boundary.  Thus, its frequency must be quantized 
according to (\ref{eq:hplusquant}); it seems that in this case
one cannot write a classical source with arbitrary 
behavior in global time.  If we simply interpret the
left-hand side of (\ref{eq:bbrel}) as a path integral over
field configurations which fall off as $(\cos\rho)^{2h_+}$,
then there will be no saddle points in the path integral when
the time dependence of the boundary configuration is arbitrary.
We leave the resolution of this conundrum for future work.

Upon imposing the conditions (\ref{eq:hplusquant}) and
(\ref{eq:hminusquant}) for $\Phi^{(+)}$ and $\Phi^{(-)}$ respectively,
the solutions may be written
in terms of Jacobi polynomials:\footnote{See Ref. \cite{erdelyi} 
for notation.}
\begin{equation}
	\Phi^{(\pm)} = e^{-i\omega t} \, Y_\ell \,  
           (\cos\rho)^{2h_\pm}
            \, (\sin\rho)^\ell \,  
	P_n^{(\ell + \frac{d}{2} - 1,2h_\pm -  \frac{d}{2})}(\cos 2\rho)\ .
\end{equation}
It is easy to show that these quantized $\Phi^{(+)}$ can be
made orthonormal
under the norm (\ref{eq:adsnorm}).  For $\nu < 1$ 
these quantized $\Phi^{(-)}$ can be made orthonormal as well.

\subsubsection{$\nu\in\IZ^{+}\cup \{ 0 \}$}
\label{sec:intmodes}

As functions of $\sin^2\rho$ the solutions $\Psi^{(1,2)}$ are the same
as before.  Because $\nu$ is the difference of the roots of the
indicial equation, however, the solutions expanded near infinity
are a little different than those listed above.
$\Phi^{(+)}$ is the same as before:
\begin{eqnarray}
	\lefteqn{\Phi^{(+)} = e^{-i\omega t} \, Y_{\ell,\{m\}}(\Omega)
		\, (\cos\rho)^{2h_+}
            \, (\sin\rho)^\ell }\nonumber\\
	&&{}_2F_1( h_+ + \half (\ell + \omega),
			h_+ + \half(\ell - \omega), 1 + \nu; \cos^2\rho )\ .
\end{eqnarray}
Again, the norm (\ref{eq:adsnorm}) is well-behaved at infinity.
For $\nu = 0$ $\Phi^{(-)}$ becomes:
\begin{eqnarray}
	\lefteqn{\Phi^{(-)} = e^{-i\omega t} \, Y_{\ell}(\Omega)
		\, (\cos\rho)^{\frac{d}{2}} 
            \, (\sin\rho)^\ell }\nonumber\\
	&&\left({}_2F_1(\frac{d}{4} + \half(\ell+\omega),\frac{d}{4}
		+ \half(\ell - \omega), 1;\cos^2\rho) 
              \, \ln\cos^2\rho \right. + \nonumber \\
	&&\ \ + \sum_{k=1}^{\infty} (\cos\rho)^{2k} \, 
		\frac{(\frac{d}{4} + \half(\ell+\omega))_k
				(\frac{d}{4} + \half(\ell-\omega))_k}{k!} \times \nonumber \\
	&&\ \ \ \ \ \left. 
		\left[h(d,0,\ell,\omega,k) - h(d,0,\ell,\omega,0)\right]\right)\ .
\label{eq:nu-eq-zero}
\end{eqnarray}
Here $h(e,f,g,\omega,k)$ is defined in (\ref{eq:hdef}). 
Note that this has a well-behaved norm at the boundary.
For $\nu> 0$ the second solution is:
\begin{eqnarray}
	\lefteqn{\Phi^{(-)} = e^{-i\omega t} \, Y_{\ell}(\Omega)
		\, (\cos\rho)^{\frac{d}{2} +\nu}
            \, (\sin\rho)^\ell }\nonumber\\
	&&\left({}_2F_1(\frac{d}{4} + \half(\nu + \ell+\omega),\frac{d}{4}
		+ \half (\nu + \ell - \omega), 1+\nu;\cos^2\rho) 
             \, \ln\cos^2\rho \right. + \nonumber \\
	&&\ \ + \left( \sum_{k=1}^{\infty} (\cos\rho)^{2k}
		\frac{(\frac{d}{4} + \half(\nu +\ell+\omega))_k
				(\frac{d}{4} + 
                  \half(\nu + \ell-\omega))_k}{(1 + \nu)_k k!} \times
					\right.\nonumber\\
	&&\ \ \ \ \ \ \ \ \ \ \left.\left(h(d,\nu,\ell,\omega,k) - h(d,\nu,\ell,\omega,0)\right)
		\right)\nonumber\\
	&&\left. \ \ - \sum_{k=1}^\nu \frac{(k-1)! (-\nu)_k}
		{(1-\frac{d}{4}-\half(\nu+\ell+\omega))_k
		(1-\frac{d}{4}-\half(\nu+\ell-\omega))_k} 
            \, (\cos\rho)^{-2k} \right) \ .
\label{eq:positive-nu}
\end{eqnarray}
The norm of these solutions blows up at the boundary.

\paragraph{Quantization condition: }

Once again we can start with the solution $\Psi^{(1)}$ which
is regular at the origin and examine its behavior at the boundary.
The transformation laws are modified when $\nu$ is integral, so
that:
\begin{eqnarray}
	\lefteqn{\Psi^{(1)} = e^{-i\omega t} Y_{\ell}(\Omega)
		(\sin\rho)^\ell \times}\nonumber\\
	&&\left(\frac{\Gamma(\nu)\Gamma(\frac{d}{2} + \ell)}
		{\Gamma(h_+ + \half(\ell+\omega))\Gamma(h_+ + \half(\ell - \omega))}
		(\cos\rho)^{2h_-} \times\right.\nonumber\\
	&&\ \ \ \ \ \left.\sum_{k=0}^{\nu-1}
		\frac{(h_- + \half(\ell +\omega))_k (h_- + \half(\ell-\omega))_k}
			{k!(1-\nu)_k} (\cos\rho)^{2k}\right. - \nonumber\\
	&&\left.\frac{(-1)^\nu\Gamma(\frac{d}{2}+\ell)}
		{\Gamma(h_- + \half(\ell+\omega))\Gamma(h_- + \half(\ell-\omega))}
		(\cos\rho)^{2h_+} \times\right.\nonumber\\
	&&\ \ \ \ \ \left.\sum_{k=0}^{\infty}
		\frac{(h_+ + \half(\ell+\omega))_k(h_++\half(\ell-\omega))_k}
			{k!(k+\nu)!} (\cos\rho)^{2k}\times\right.\nonumber\\
	&&\ \ \ \ \ \ \ \ \ \ \left.\left[\ln\cos^{2}\rho - \psi(k+1) - \psi(\nu+k+1)\right.\right.
		\nonumber\\
	&&\ \ \ \ \ \ \ \ \ \ \left.\left.
		+\psi(h_+ + \half(\ell+\omega)+k)+ \psi(h_+ + \half(\ell-\omega)+k)
		\right] \right)\ .
\label{eq:int-nu-psi}
\end{eqnarray}
Once again, in order to isolate normalizable modes which fall off as
$(\cos\rho)^{2h_+}$ at the boundary we must impose (\ref{eq:hplusquant}).
At these frequencies, the gamma functions in the denominator of the
coefficient of the final sum have negative integer argument; so only terms
in the sum which have compensating poles will survive.  Such poles will
come from one of the final two $\psi$ functions; thus, the logarithmic term
drops out and the solution falls off at the boundary as desired, giving a
series of modes built from $\Phi^{(+)}$.  Equivalently, one can easily
show that when (\ref{eq:hplusquant}) holds, $\Phi^{(+)}$ is regular at the
origin.  As before, for general $\omega$ we have non-fluctuating modes
which are well-behaved at the origin and which have a logarithmic part at
infinity.  A particularly interesting set of non-normalizable solutions
occurs when
\begin{equation}
	\omega = 2h_- + \ell + 2n;\ \ \ \ \ n = 0,1,\ldots \nu-1\ .
\label{eq:intnonnorm}
\end{equation}
For such frequencies the final sum in (\ref{eq:int-nu-psi}) vanishes
and the result is, as in the case of non-integral $\nu$, a rational function in
$\cos\rho$.  We will see in the next section that these solutions are part
of a special highest weight representation that exists for integral $\nu$.

\section{$\ads{3}$ and $\SL\times\SL$}
\label{sec:ads3}

The zoo of solutions that we have described in global coordinates  
should fall in various representations
of the spacetime isometry/boundary conformal group.  The fluctuating
modes should clearly fall in unitary representations, and the boundary
operators should create states (in sectors for which the state-operator map is
one-to-one)  which fall in such unitary representations as well.  The
non-normalizable modes do not have to fall in unitary representations
of the conformal group,
but we will see that they do lie in linear representations.  Since they
couple to primary boundary operators and their conformal descendants, 
such representations are also important; in order to understand the coupling
of descendants we need to understand how the various representations
combine.

We will discuss mode solutions in $\ads{3}$,
for which the representations of the isometry group are well-known;
see especially \cite{barut}\ for a discussion and references.
The highest-weight unitary representations in global coordinates and the
continuous representations in Poincar\'e coordinates were discussed
in \cite{jaex}.  As we will see, the results of the previous section
provide explicit expressions for the wavefunctions for all of the
linear representations.\footnote{For previous discussions of
$\SL\times\SL$ structure in solutions to the wave
equation in black hole backgrounds, see the
review \cite{cvetic-larsen} and the references therein.}

As discussed in Appendix~\ref{app:3coords}, $\ads{3}$ is
obtained as the hyperboloid $-\cosm^2 = -U^2 -V^2 + X^2 +Y^2$ embedded in
$\CR^{2,2}$ with metric $ds^2 = -dU^2 -dV^2 +dX^2 +dY^2$.  The isometry
group of $\CR^{2,2}$ is clearly $SO(2,2)$ generated by:
\begin{eqnarray}
J_{01} &=& V\partial_U - U \partial_V
~~~~~;~~~~~
J_{02} = X\partial_V + V \partial_X
~~~~~;~~~~~
J_{03} = Y\partial_V + V \partial_Y 
\nonumber \\
J_{23} &=& X\partial_Y - Y \partial_X
~~~~~;~~~~~
J_{12} = X\partial_U + U \partial_X
~~~~~;~~~~~
J_{13} = Y\partial_U + U \partial_Y 
\label{eq:4dgens}
\end{eqnarray}
We can construct two commuting $\SL$ factors from these generators as
$\SL_L= \{L_1 =(J_{01} + J_{23})/2, L_2 =(J_{02} - J_{13})/2, L_3
=(J_{12} + J_{03})/2 \} $, and
$\SL_R = \{\tL_1 =(J_{01} - J_{23})/2, \tL_2 = (J_{02} + J_{13})/2, \tL_3
=(J_{12} - J_{03})/2 \} $.   With these definitions:
\begin{equation}
[L_1,L_2] = - L_3 ~~~~~;~~~~~ [L_1,L_3] = L_2 ~~~~~;~~~~~
[L_2, L_3] = L_1
\end{equation}
and similarly for the $\tL$.   These generators preserve
the hyperboloid that embeds $\ads{3}$ in $\CR^{2,2}$ and so  are also
isometries of $\ads{3}$.

 \mbox{} From $\{L_1,L_2,L_3\}$ it is easy to construct linear combinations
$\{L_0,L_\pm\}$ that satisfy the algebra:
\begin{equation}
[ L_0, L_\pm ] = \mp L_\pm ~~~~~;~~~~~ [L_+,L_-] = 2 L_0
\end{equation}
The quadratic Casimir of  $\SL_L$ is then:
\begin{equation}
L^2 = {1\over 2} (L_+ L_- + L_- L_+) - L_0^2
\label{eq:casdef}
\end{equation}
Representations can be built by starting with a state $|\psi\rangle$
with $L_0$ eigenvalue $E$ and acting on it with powers of $L_+$ or $L_-$.
The commutation relations as defined imply that $L_-$ ($L_+$)
raises (lowers) the $L_0$ eigenvalue by one unit.  Highest (lowest) weight
representations contain a state $|h\rangle$ that is annihilated
by $L_+$ ($L_-$).  The entire representation can be built by acting on this
state with arbitrary powers of $L_-$ ($L_+$).

\subsection{Review of the representations of $\SL$.}

The irreducible representations of the $\SL$ algebra are well known
\cite{bargmann}.  Barut and Fronsdal \cite{barut}\ derived a set of linear
representations which contain them; we will follow their discussion as we
will find that this more general set is important for our purposes.
~\footnote{We use generators with a slightly different normalization than
\cite{barut}.  The generators in \cite{barut} are called $L_{12}$ and
$M^{\pm}$, and the Casimir is called $Q$.  In this notation our generators
are $L_0 = L_{12}$ and $L_\pm = i\sqrt{2}M^{\mp}$; our expression for the
Casimir is $L^2 = - Q$.}  The representations are indexed by 2 invariants.
The first is $h$ (called $-\Phi$ in \cite{barut}) which is related to the
quadratic Casimir:
\begin{equation}
	L^2 = h(h-1)\ .
\label{eq:casimir}
\end{equation}
We will see that this is the same $h$ as defined in the previous section;
$h_\pm$ are the two solutions for a given value of the Casimir.  The second
invariant is the fractional part $E_0$ of the spectrum of $L_0$ for a given
representation (here
we use the same notation as in \cite{barut}.)  Representations are filled out
by starting with a given vector in the representation and acting on it an arbitrary
number of times with $L_\pm$.  The resulting representations are:
\begin{itemize}
\item $\CD(L^2, E_0)$.  This an irreducible,
infinite-dimensional representation but
does not have a highest or lowest weight state. $h$ and $E_0$ are not
related; the only condition is that $h\pm E_0$ is not an integer.  $E$
does not even have to be real, but since we wish to describe
stable modes in spacetime we will not consider complex values.
We can define $-\half < E_0 \leq \half$ without losing generality;
the spectrum is $L_0 = E_0 + n$ for $n$ an arbitrary integer.
For fixed $L^2$ and $E_0$,
the representations for each branch of  (\ref{eq:casimir})\ are equivalent. 
The non-unitary representations will correspond to non-normalizable modes
for which the energies are not related to the mass, i.e. for which $\omega$
is not quantized in even integers above $2h_- + \ell$.
Imposing unitarity restricts $\CD(L^2, E_0)$ to two types of representations:
\begin{enumerate}
	\item $\CD_P$ -- the ``principal series'' occurs for $L^2 < -\frac{1}{4}$;
	thus $h = \half + i\lambda$.
	$\lambda \neq 0$ will correspond to unstable modes in spacetime
	as noted in \cite{brfreed,holowit}.
	\item $\CD_S$ -- the ``supplementary series'' occurs
	for $L^2 > - \frac{1}{4}$.  Here $h$ is real and
	$|h - \half| < \half - |E_0|$.  This occurs in the range
	$0 < \nu < 1$ discussed in the previous section.
\end{enumerate}
\item $\CD^+(h,E_0)$.  This is an irreducible, infinite-dimensional highest
weight representation and exists for $2h \notin \IZ^-\cup{0}$.  $E_0=h$ and
the spectrum is $L_0 = E_0 + n$ for integral $n \geq 0$; the highest weight
$n=0$ state is annihilated by $L_+$.  The representation will be unitary if
$h > 0$.  These states correspond to the solutions in global coordinates
quantized according to (\ref{eq:hplusquant}) for general $\nu$.  For
non-integral $\nu$ solutions quantized according to (\ref{eq:hminusquant})
also transform in this representation.  The solutions $\Phi^{(+)}$ have
$h=h_+>0$ and live in a unitary positive energy
representation. The solutions $\Phi^{(-)}$ have $h=h_-$; 
they live in a non-unitary representation for
$\nu > 1$ and in a unitary representation for $\nu < 1$.  The derivation of this
representation in \cite{barut}\ shows that it can be imbedded in a
reducible, nondecomposable representation where $n$ is an arbitrary
integer.  In this representation we can reach $\CD^+$ by starting with
negative $n$ states and acting on them repeatedly with $L_-$; however, once
we examine states in $\CD^+$ we cannot leave the irrep with actions of
$L_\pm$ because of the highest weight state.
\item $\CD^-(h,E_0)$.  This is an irreducible,  infinite-dimensional
lowest weight representation
and again exists for $2h \notin \IZ^- \cup\{0\}$.  $E_0 = -h$
and the representation is unitary for $h > 0$. The spectrum is
$L_0 = E_0 - n$ for integral $n \geq 0$.  The
lowest weight $n=0$ state is annihilated by $L_-$.  These are the
negative energy modes corresponding to $\CD^+$.  Ref. \cite{barut}\
imbeds this in a reducible, nondecomposable representation which contains
energies larger than that of the lowest weight (highest energy) state.
\item $\CD(h)$.  This is an irreducible, finite-dimensional representation.
It occurs when $2h \in \IZ^-\cup\{0\}$ and $E_0 = 0$.  Its
spectrum is $L_0 = h + n$ for integral $0 \leq n \leq -2h$.  The
representation is only unitary in the case $h = 0$, i.e. for the 
identity representation, 
also known as the singleton~\cite{singleton,fronsdal,single-three}.
It is contained
in a reducible nondecomposable representation for which $n$ is
arbitrary.  $\CD(h)$ arises in $\ads{3}$ as $\Phi^{(-)}$ for
integral $\nu$.  In global coordinates, the case $\ell = 0$,
$\omega = 2 h_- \cdots -2 h_-$ for integral $\omega$ corresponds to
a tensor product $\CD(h_-) \times \CD(h_-)$ transforming
under $\SL_L \times \SL_R$.  For arbitrary $\ell$
nondecomposable representations
containing this irrep can occur.  
\end{itemize}
In the next subsection we will explicitly discuss how these representations
are realized as solutions to the wave equation in $\ads{3}$ in global coordinates.

\subsection{Global coordinates}
In global coordinates the $\ads{3}$ metric is:
\begin{equation}
ds^2 = \cosm^2 [ - \cosh^2\mu \, dt^2 + d\mu^2 + 
\sinh^2\mu \, d\theta^2]
\label{eq:metglob}
\end{equation}
and a scalar field of mass $m$ has a wave equation $(\Box - m^2\cosm^2)\phi
= 0$ where:
\begin{equation}
\Box = \partial_\mu^2 + {2\cosh(2\mu) \over \sinh(2\mu)} \partial_\mu
+ {1 \over \sinh^2\mu} \partial_\theta^2 - {1 \over \cosh^2\mu}
\partial_t^2
\label{eq:globwave}
\end{equation}
In these coordinates, a convenient basis for $\SL_L$ is:
\begin{equation}
L_0 = iL_1 ~~~~~;~~~~~ L_+ = (L_2 + i L_3) ~~~~~;~~~~~ 
L_- = -(L_2 - iL_3) 
\label{eq:globbasis}
\end{equation}
Starting with generators in (\ref{eq:4dgens}) and using the
coordinate patch in Appendix~\ref{app:3coords} that yields the global
metric, it is easy to work out the explicit representation:
\begin{eqnarray}
L_0 &=& i\partial_w 
\label{eq:glob0}\\
L_- &=& i e^{-iw} \left[ {\cosh(2\mu) \over \sinh(2\mu)} \partial_w -
                          {1 \over \sinh(2\mu)} \partial_{\bw} +
                          {i \over 2} \partial_\mu \right] 
\label{eq:glob-}\\
L_+ &=& i e^{iw} \left[ {\cosh(2\mu) \over \sinh(2\mu)} \partial_w -
                          {1 \over \sinh(2\mu)} \partial_{\bw} -
                          {i \over 2} \partial_\mu \right] 
\label{eq:glob+}
\end{eqnarray}
where $w = t+\theta$ and $\bw=t-\theta$. The generators $\bar{L}_0$,
$\bar{L}_\pm$ of $\SL_R$ are
obtained by exchanging $w$ and $\bw$
in (\ref{eq:glob0}--\ref{eq:glob+}).  Clearly $L_0 \pm \tL_0$
generate time translations and rotations. 

\paragraph{Discussion of explicit solutions: } The d'Alembertian for
scalar fields is given in terms of the left and right Casimirs as:
\begin{equation}
\Box \Phi = -2(L_L^2 + \tL_R^2) \Phi = m^2 \cosm^2\ .
\label{eq:caswave}
\end{equation}
The highest weight states $\CD^{(+)}$ and $\CD$ are simplest to describe.
We require that $L_+ \Phi_H =
\tL_+ \Phi_H = 0$, and this imposes $h=\hdim$.   Using this, the
d'Alembertian in (\ref{eq:caswave}) acting on highest weight states
reduces to:
\begin{equation}
h(h-1) = {m^2\cosm^2 \over 4}  ~~~~~  \Longrightarrow ~~~~~
h_\pm = {1\over 2}( 1 \pm \sqrt{1 + m^2 \cosm^2} )
\end{equation}
Explicit solutions of the equation $L_+ \Phi_H = \tL_+ \Phi_H = 0$ give
\cite{jaex}:
\begin{equation}
\Phi_H^{(\pm)} = e^{-ih_\pm w -ih_\pm \bw} {1\over (\cosh\mu)^{2h_\pm}} =
e^{-i(2h_\pm) t} {1\over (\cosh\mu)^{2h_\pm}}
\end{equation}
so that $h$ in this section and in the previous section
are the same.  

For $\nu = h_+ - h_-$ non-integral, $\Phi^{(+)}_H$ are precisely the
minimum-energy normalizable modes found in Sec.~\ref{sec:globsoln} (here
$\tan\rho = \sinh\mu$ as in Appendix~\ref{app:3coords}).  $\Phi^{(-)}_H$
are the non-normalizable modes of lowest energy in the spectrum
(\ref{eq:nonnormquant1}).  Other non-fluctuating modes will live in
non-highest-weight representations.  Descendant states are constructed on
the primary $\Phi_H^{(\pm)}$ by the action of $(L_-)^p (\tL_-)^q$ and have
weights $h= h_\pm + p$ and $\hdim = h_\pm + q$.  Examining the differential
operator $L_-$ shows that all of these solutions have the same boundary
behaviour as the primary states and therefore share their normalizability
properties.  Finally, $L_0 + \tL_0$ is the generator of time translations,
and $L_0 - \tL_0$ is the generator of rotations, so that the frequency and
angular momentum are given by $\omega = h + \hdim$ and $\ell = h - \hdim$.
The spectra of the two towers of states are given by:
\begin{equation}
\omega_\pm = 2n + 2h_\pm + \ell;~~~~~ n=0,1,2,\ldots
\end{equation}
This matches the spectra for the $\nu\notin\IZ$ 
solutions found in Sec.~\ref{sec:globsoln}.

For integral $\nu$, the situation is the same for 
representations built on $h_+$; 
the modes living in this representation
are again the normalizable solutions $\Phi^{(+)}$.  
The non-fluctuating states will fall into two types.
If $\omega$ is not quantized according to (\ref{eq:intnonnorm}),
the solutions will live in the representations $\CD(L^2,\omega)$.
If the energies are quantized according to (\ref{eq:intnonnorm}),
then the solutions $\Phi^{(-)}$ will live
in nondecomposable representations
containing the finite-dimensional representations $\CR = \CD(h_-)\times
\CD(h_-)$. 
The lowest energy (highest weight)
state in $\CR$ has $\omega = 2h_-$ and is given by
\begin{equation}
	e^{-iw h_- - i \bw h_-} (\cosh\mu)^{-2h_-}\ .
\end{equation}
The highest energy (lowest weight) state in $\CR$ has $\omega = 
-2h_- = 2h_+ - 2$ and is given by
\begin{equation}
	e^{iw h_- + i \bw h_-} (\cosh\mu)^{-2h_-}
\label{eq:lowest-weight}
\end{equation}
which is annihilated by $L_-$ and $\tL_-$.
One may also find these modes in the manner 
described in Sec.~\ref{sec:intmodes}.

The case $\nu = 1$ is quite straightforward and interesting.  
In addition to the (unitary) identity representation 
at $\ell = 0$ there are solutions with
$(h \in \IZ^+, \bar{h} = 0)$:
\begin{equation}
	\Phi = e^{-i\ell w}\tanh^\ell \mu \ .
\end{equation} 
One may apply $L_\pm$ to reach
other such solutions.  For general $\ell$, applying
$L_\pm$ to $(h,\bar{h}) = (\ell, 0)$ gives us the
state $(h,\bar{h}) = (\ell\mp 1,0)$.   
Applying $L_+$ to $(h=1,\bar{h}=0)$
gives the irreducible identity representation $\CD(0)\times\CD(0)$.
Applying $\bar{L}_+$ annihilates all states $(h,\bar{h})=(\ell,0)$. 
Applying $\bar{L}_-$ to such states
brings one to the normalizable set of solutions
forming the irrep
$\CD^+(1,0)\times\CD^+(1,0)$.  There are similar
solutions for $(h=0, \bar{h} \in \IZ^+)$ and
states generated from these by applying the raising
and lowering operators.  The full representation
in this example is a nondecomposable
representation and was discussed in Ref. \cite{single-three}.  
Such representations were discussed for
$\nu = 0,1$ in $\ads{d+1}$ in refs \cite{ferrara-one, ferrara-two}\
from a somewhat different point of view.

In summary, we have found that every solution to the wave equation which is
well-behaved in the bulk lies in a representation of the bulk
isometry/boundary conformal group $G=\SL_L\times\SL_R$.  The energy is
simply the $L_0$ eigenvalue, and the singularity or zero of the
wavefunction at infinity is related to the invariant $h$ of the
representation.  The map is quite natural.  Normalizable modes (plus the
singleton) correspond to unitary representations of $G$; it makes sense to
quantize these modes.  Non-normalizable modes correspond to non-unitary
representations of $G$.  These are the modes we wish to keep as
non-fluctuating backgrounds.

\paragraph{Relation to Conformal Field Theory on The Cylinder:}
As $\mu \rightarrow \infty$, the generators $\SL_L$ acting on
surfaces of fixed $\mu$ become:
\begin{equation}
L_0 = i \partial_w ~~~~~;~~~~~
L_- = i e^{-iw} \partial_w ~~~~;~~~~
L_+ = i e^{iw} \partial_w
\end{equation}
These generators and their $\SL_R$ companions which replace $w$ by $\bw$
are the standard conformal symmetries of the cylinder.   Acting on
descendants of a primary state $\Phi_H$ with weight $h$, the 
generators at the boundary become:
\begin{equation}
L_0 = i \partial_w ~~~~~;~~~~~
L_- = i e^{-iw} (\partial_w - ih) ~~~~;~~~~
L_+ = i e^{iw} (\partial_w + ih)
\end{equation}
The shift of $L_\pm$ by $\pm ih$ arises from the radial derivatives 
in the bulk generators.

\subsection{Poincar\'e Coordinates}
In Poincar\'e coordinates the $\ads{3}$ metric and d'Alembertian are:
\begin{eqnarray}
ds^2 &=& \left({\cosm^2 \over r^2 } \right) (-dt^2 + dx^2 + dr^2) \\
\Box &=& r^2 \, \partial_r - r \, \partial_r + 
r^2 (\partial_x^2 - \partial_t^2) \ .
\end{eqnarray}
  As discussed in Appendix~\ref{app:3coords} these coordinates only cover a
region of $\ads{3}$ and there is a horizon at $r = \infty$.  A
convenient basis for $\SL_L$ in these coordinates is:
\begin{equation}
L_0 = -L_2 ~~~~~;~~~~~ L_+ = i(L_1 + L_3) ~~~~~;~~~~~ L_- = i(L_1 - L_3)
\label{eq:plin}
\end{equation}
Starting with the explicit generators in (\ref{eq:4dgens}) and
implementing the Poincar\'e coordinates in Appendix~\ref{app:3coords}
yields the generators:
\begin{eqnarray}
L_0 &=& {-r \over 2} \, \partial_r - z \, \partial_z \label{eq:P0}\\
L_- &=&  i  \cosm \, \partial_z \label{eq:P-}\\
L_+ &=& -{i \over \cosm} \left[ z r  \, \partial_r + z^2 
             \, \partial_z + r^2  \, \partial_{\bz} \right]
            \label{eq:P+}
\end{eqnarray}
Here $z = t + x$ and $\bz = t - x$.  The generators $\tL$ of
$\SL_R$  simply exchange $z$ and $\bz$.

\paragraph{Translations and CFT on the plane: }
It is instructiive to examine the action of the generators
(\ref{eq:P0}) -- (\ref{eq:P+}) on surfaces of constant $r$ at
the boundary $r=0$:
\begin{equation}
L_0 = -z\, \partial_z   ~~~~~;~~~~~
L_- = i \cosm \, \partial_z ~~~~~;~~~~~
L_+ = {-i z^2 \over \cosm} \, \partial_z 
\end{equation}
These are the standard generators of conformal transformations on the
plane.  (The factors of $i$ and $\cosm$ arise because we are in Minkowski
space and $z$ is a dimensionful coordinate.)  In the case of
the CFT on the cylinder, $L_0$ generated 
translations. 
$L_-$ generates translations on the plane,
while $L_0$ generates dilatations.  Furthermore, the basis
(\ref{eq:plin}) for $\SL$ is different from the basis (\ref{eq:globbasis})
we used for global coordinates.
As discussed in the Appendix,
the boundary of the patch of spacetime covered by Poincar\'e
coordinates is conformal to Minkowski space and we have chosen 
the corresponding natural basis for $L_0$ and $L_\pm$.   It is
important to emphasize that we are {\em not} dealing with the standard
bijective map between CFTs on the cylinder and the plane.  The plane 
that  appears at the Poincar\'e  boundary is merely a patch of the 
cylinder and can only be expected to describe the theory on the
cylinder in a thermal sense, after tracing over some degrees of
freedom. 

The Poincar\'e mode solutions of Sec.\ref{sec:poinsoln} with fixed
frequency $\omega$ and momentum $k$ are eigenstates of $L_-$ and $\tL_-$
with eigenvalues $\omega \pm k$ (see also \cite{jaex}).  
$L_-$ generates a non-compact subgroup of $\SL$ \cite{barut};
thus the wavefunctions we get by diagonalizing this operator
should be continuously moded, as we have found.
 
\paragraph{Representing the isometries: }
We can also construct a set of scalar fields in Poincar\'e
coordinates that carry well-defined weights under $L_0$ and $\tL_0$.
Again, the d'Alembertian is the sum of left
and right Casimirs:
\begin{equation}
\Box \Phi = -2(L_L^2 + \tL_R^2) \Phi = m^2 \cosm^2
\end{equation}
where we have used the definition in (\ref{eq:casdef}) of $\SL$
Casimirs. States of weight $(h,\hdim)$ under $L_0$ and $\tL_0$ must
satify the equations:
\begin{equation}
-\left({r \over 2} \, \partial_r + z \, \partial_z \right)
\Phi = h \, \Phi 
~~~~~;~~~~~
-\left({r \over 2} \, \partial_r + \bz \, \partial_{\bz} \right)
\Phi = \hdim \, \Phi 
\end{equation}
Since $L_0$ and $\tL_0$ are linear in derivatives, a product of
eigenstates of these operators is still an eigenfunction.   
A general class of eigenstates is given by:
\begin{equation}
\Phi = r^a \, z^{b} \, \bz^{c}
       (r + \sqrt{z\bz})^d 
       (r - \sqrt{z\bz})^e 
\label{eq:phip}
\end{equation}
The left and right weights are then:
\begin{equation}
(h,\hdim) = \left( -[b + (a+d+e)/2], -[c + (a + d + e)/2] \right)
\end{equation}
In global coordinates the eigenstates of $L_0$ essentially provided a
Fourier basis for mode expansions.  Here the $L_0$ eigenstates provide an
expansion in a power series of functions.

To find highest weight (primary) states we want to additionally solve the
equations $L_+ \Phi_H = \tL_+ \Phi_H = 0$.  Simultaneous solution of these
conditions requires that $z\partial_z \Phi = \bz \partial_{\bz} \Phi$
implying that $h = \hdim$ and 
imposing symmetry between $z$ and $\bz$ in primary states.
Using $z\partial_z \Phi = \bz \partial_{\bz} \Phi$ and the equation for
$L_0 \Phi = h \Phi$ in $L_+ \Phi_H = 0$ gives:
\begin{equation}
\partial_r \Phi_H = {-2h \over r} \, 
\left( {r^2 + z\bz \over r^2 - z\bz} \right) \, \Phi_H
\end{equation}
which is easily solved to give:
\begin{equation}
\Phi_H = r^{2h} (r^2 - z\bz)^{-2h}
\label{eq:phighest}
\end{equation}
It can be explicitly checked that this is a primary state with weights
$(h,h)$.  Requiring that the d'Alembertian acting on this solution has
eigenvalue $m^2\cosm^2$ yields the condition:
\begin{equation}
h(h-1) = {m^2\cosm^2 \over 4}  ~~~~~  \Longrightarrow ~~~~~
h_\pm ={1\over 2}( 1 \pm \sqrt{1 + m^2 \cosm^2} )
\end{equation}
exactly as in the case of global coordinates.  

For $m^2 > 0$ and $z\bz < 0$ the solutions with $h= h_+$ vanish at the
boundary ($r =0$) and at the horizon ($r =\infty$) while the the $h_-$
solutions diverge in both locations.  This agrees with
the claim in Sec.~\ref{sec:norm} that the $h_-$ representation
alters the boundary conditions but does not fluctuate.  Furthermore, it suggests that in
Poincar\'e coordinates the horizon should also be treated as a boundary
with which flux can be exchanged.  Indeed, it is well known that
quantum field theory in a spacetime with horizon requires the specification
of horizon boundary conditions.  For $z \bz \geq 0$ the situation is more
disturbing - the $h_+$ solution is singular in the bulk of spacetime at
$r^2 = z\bz$.  This appears to be a pathology that arises because the
surface $(r^2 - z\bz) = 0$ is a fixed point of $L_0$, the generator of
dilatations.  The eigenstates of $L_0$ are accordingly singular on this
surface. 

\section{Discussion and conclusions}

\subsection{Understanding the bulk from the boundary}
\label{sec:silliness}

The original motivation for this work was the desire to study 
quantum gravity in the bulk spacetime from the perspective
of the boundary gauge theory.   In particular, we would like to study the
appearance of spacetime singularities and horizons
(see~\cite{garyhirosi,garysimon}, for example) which we expect to
be related to nonperturbative issues in the gauge theory.   A 
preliminary step is to determine whether we can say anything about
local spacetime physics from the boundary perspective.

The existence of the normalizable modes displayed in this article  
implies that there is a natural Hilbert space of small
fluctuations around the bulk AdS background.  These fluctuating modes
are the probes of bulk causal structure. Since we can map such states 
into the boundary Hilbert space, we might expect that there is an 
analysis of local bulk processes from the boundary point of view.

On the other hand it seems that we cannot reconstruct position 
space correlation functions in terms of the correspondence as written in
(\ref{eq:bbrel}).   To see this let us expand both sides of
(\ref{eq:bbrel}) in a formal series in the boundary field
$\Phi_b$.  For the moment we will work in Euclidean space where 
this is a well-defined procedure since the bulk field $\Phi$ is 
uniquely determined by $\Phi_b$.  (We will return to the Lorentzian 
version below.)   As pointed out in~\cite{holowit}, 
$\Phi$ can be expressed in terms of $\Phi_b$ via the equation
(here, in Poincar\'e coordinates):
\begin{equation}
        \Phi (x^0, \vec{x}) = \int_{\CB} d\vec{x}'
                K(x^0, \vec{x}; \vec{x}') \Phi_b (\vec{x}')\ .
\label{solution}
\end{equation}
where $K$ is a solution to the wave equation behaving as a delta
function times a given power of $x^0$ on the boundary.  The quadratic piece of
the action can be written as:
\begin{equation}
        S_{{\rm eff}} = \int dz dx dz' dx' \Phi(z,x) \CF (z,x;z',x')
                \Phi(z',x') + \ldots
\end{equation}
$\CF$ is the inverse spacetime propagator for $\Phi$, which
we would like to extract
from the boundary theory.  This piece can be written as a 
quadratic expression in $\Phi_b$ via  (\ref{solution}).
Expanding (formally) (\ref{eq:bbrel}) to quadratic order, we find:
\begin{eqnarray}
        \langle \CO(t_1,\vec{x}_1)\CO(t_2,\vec{x}_2) \rangle =
                -2 i \int dt'd\vec{x}' dr' dt'' d\vec{x}'' dr''
                K(r',t',\vec{x}'; t_1,\vec{x}_1) \times \\
                \ \ \times\CF(r', t', \vec{x}'; r'', t'', \vec{x}'')
                K(r'', t'', \vec{x}''; t_2, \vec{x}_2 )\ .
\label{eq:twopoint}
\end{eqnarray}
Extracting $\CF$ from this would require ``inverting'' $K$, which seems
impossible.  The integration over $r',r''$ washes out any information
about localization in the direction perpendicular to the boundary.
Essentially, this point was made in ref.~\cite{banksgreen} -- the
correspondence (\ref{eq:bbrel}) relates off-shell operators in the boundary
to on-shell fields in the bulk.  Since the $r$ dependence of the latter
depends on the $(t,\vec{x})$ dependence, full localization is not possible. 

The point is that instead of being able to reconstruct arbitrary 
off-shell correlators, we
must be content with a description of on-shell quantities in the bulk.
Indeed, our knowledge of the mapping between the bulk and boundary  
Hilbert spaces and their Hamiltonians implies that transition amplitudes
between arbitrary {\em physical} states are computable in either of the
dual  descriptions. 
Off-shell physics can be probed to the extent that on-shell correlators
receive contributions from off-shell modes in intermediate states.  This
situation will force the boundary analysis of the bulk geometry to 
be somewhat subtle and indirect, but still possible in principle.  

In addition to correlation functions,  we want to be able to
ask how the vacuum in the bulk string theory relates to the boundary
field theory.  This question really concerns the
global causal structure -- the existence of different natural vacua in
a spacetime often reflects the presence of horizons, for example.
Understanding this is also relevant to the study of
possible singularities 
in black hole backgrounds from the dual gauge theory point of view.
It may be that although local bulk physics is
quite difficult to examine via the boundary field theory, 
the global causal structure may somewhat
easier to address.

\subsection{Apparent ambiguities in the effective action}

The presence of normalizable modes in Lorentzian AdS might appear to 
render the correspondence (\ref{eq:bbrel}) ambiguous,
since there is no preferred solution corresponding to a given 
boundary value. 
However, there is a natural prescription - we must
sum the effective action over all
spacetime backgrounds with the same boundary behavior, with a weighting
which depends upon the state of the system.  
We can generalize
the calculations in \cite{holowit}\
that we outlined above, by appropriately modifying
(\ref{solution}):
\begin{equation}
        \Phi  = \Phi_n ~+~\int_{\CB}
                K \Phi_b = \Phi_n + \Phi_{nn} \ ,
\end{equation}
where $\Phi_n$ is an arbitrary normalizable solution of the field
equations.  Here $K$ is a particular solution we have chosen 
which solves the wave equation and has
the same behaviour at infinity as in the Euclidean case.
$S_{{\rm eff}}$ can be written as:
\begin{equation}
	S_{{\rm eff}}= \int \left[\partial (\Phi_n + \Phi_{nn})\right]^2 + m^2 \Phi^2\ .
\label{msolution}
\end{equation}
Upon integrating by parts, the bulk term vanishes.  It is easy to
see that upon including the correct measure factors, the
non-vanishing boundary terms are:
\begin{equation}
	S_{\CB} = \int_\CB dS^i\left[ 2\Phi_n\p_i \Phi_{nn} + \Phi_{nn} \p_i \Phi_{nn}
		\right]\ .
\end{equation}  
When summing over field histories in the Lorentzian path
integral, one must also specify a state at early and late times in the
form of wavefunctionals $\Psi_{i,f}[\Phi]$.  A path integral obtained by
continuation from Euclidean signature will  pick out vacuum wavefunctionals,
since the $i\epsilon$ prescription damps out all excited states.  In such a
case, the normalizable mode $\Phi_n$ is set to zero and the boundary
contribution to the action in (\ref{msolution}) vanishes.   More generally,
one could choose to study excited states at early and late times by 
explicitly including the appropriate wavefunctionals; in this case the 
boundary terms become physically relevant and give contributions to 
correlation functions evaluated in excited states.


For  calculations in interacting theories, the
effects of the ambiguity can be more pronounced.   In Euclidean
AdS, a dramatic example of the effect of multiple saddlepoints 
with same boundary behaviour is
the high-temperature transition between AdS space and the AdS black hole
discussed in Sec. (3.2) of~\cite{holowit}.   Our normalizable modes
represent a large class of saddlepoints of the bulk action in 
Lorentzian AdS and, in an interacting theory, they encode the 
non-trivial dynamics of the bulk.



\subsection{Conclusions}

In this work we have developed a Lorentzian signature version of the 
bulk-boundary correspondence.  This required understanding the respective
roles played by normalizable and non-normalizable modes.  The two sets
of modes emerge naturally, either from direct solution of the field 
equations or from the field representations of the AdS
isometry group.  The non-normalizable modes act as backgrounds and couple
to local operators in the boundary description, while normalizable modes
describe fluctuations in the bulk. 

The picture we have presented suggests several avenues for the study
of black hole spacetimes from the boundary perspective.  Black holes
can be constructed in AdS spaces \cite{BTZ} by making discrete identifications
of the geometry.   Unlike pure AdS, the resulting bulk spacetime has 
global horizons and singularities in the classical supergravity 
approximation.  The question is whether and how the boundary theory
describes the interior of the black hole.  One would hope
that the Hilbert space of states within the black hole is identified
with a sector of states in the boundary theory; this would 
realize a form of
the black hole complementarity advocated by 't Hooft and Susskind.   
In our picture, this issue would be studied by considering the
roles of the normalizable and non-normalizable solutions to the 
wave equation in the black hole spacetime.

\vspace{0.2in} {\bf Acknowledgments:} We have enjoyed discussions with
E. Gimon, S. Kachru, E. Keski-Vakkuri, R. Leigh, E. Martinec, S. Mathur,
J. Schwarz, E. Silverstein, A. Strominger and C. Vafa.  We are also
grateful to O. Aharony, J. Maldacena, R. Helling, R. Myers, and S. Ross for
useful comments on the original preprint version of this work.  P.K. thanks
Stanford University for their hospitality during the course of this work.
V.B. is supported by the Harvard Society of Fellows and by the NSF under
grant NSF-Phy-92-18167.  P.K. is supported by in part by DOE grant
DE-FG03-92-ER40701 and by the DuBridge foundation.  A.L. is supported by
NSF grant NSF-Phy-92-18167.

\appendix
\section{Coordinate systems on $\ads{d+1}$}
\label{app:coords}

\subsection{$\ads{3}$}
\label{app:3coords}
$\ads{3}$ is defined as the hyperboloid $-U^2 -V^2 + X^2 +Y^2 = -\cosm^2$
embedded in a space with metric $ds^2 =-dU^2 - dV^2 +dX^2 + dY^2$.   

\paragraph{Global coordinates: } Global coordinates for $\ads{3}$ are defined
by:
\begin{eqnarray}
U &=& \cosm \cosh\mu \sin t ~~~~~~;~~~~~~ V = \cosm \cosh\mu \cos t
\nonumber \\
X &=& \cosm \sinh\mu \cos\theta ~~~~~~;~~~~~~ Y = \cosm \sinh\mu \sin\theta 
\nonumber
\end{eqnarray}
These yield the metric:
\begin{equation}
ds^2 = \cosm^2 [ - \cosh^2\mu \, dt^2 + d\mu^2 + 
\sinh^2\mu \, d\theta^2]
\label{eq:globalmet}
\end{equation}
Here $0 \leq \mu \leq \infty$, $0\leq \theta \leq 2\pi$ and $0 \leq t
\leq 2\pi$.  We unwrap $t$ to have range $-\infty$ to $\infty$ in order to
work on ${\rm CAdS}_3$, the universal cover of $\ads{3}$.  

It is often convenient to make the coordinate tranformation $\sinh\mu =
\tan\rho$ with $0\leq \rho \leq \pi/2$.  The metric then becomes:
\begin{equation}
ds^2 = \cosm^2 [-\sec^2\rho \, dt^2 + \sec^2\rho \, d\rho^2 + \tan^2\rho \,
d\theta^2] 
\label{newglobal}
\end{equation}

\begin{figure}                                 
\begin{center}                                 
\leavevmode                                 
\epsfxsize=2.8in
\epsfysize=2.5in                                 
\epsffile{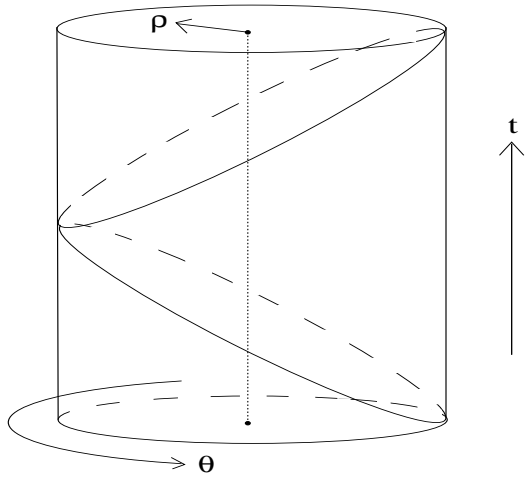}                                 
\end{center}                                 
\caption{Anti-de Sitter spacetime displayed as the interior of a
cylinder.  For the single cover of AdS the top and bottom boundaries should
be identified, whereas for its universal covering space (CAdS) an infinite
number of copies should be attached above and below the displayed region.
The boundary of AdS is identified with the boundary of the cylinder.
The coordinates indicated correspond to those in (89).
Horizons in AdS are obtained by making two diagonal cuts through the
cylinder, as shown.  The cuts divide AdS into two regions, each of which is
covered by a set of Poincare coordinates.  The boundary divides into two
diamond shaped regions, which are each conformal to copies of flat
Minkowski space.
\label{fig:one}}
\end{figure}

\mbox{} From the above we see that $\ads{3}$ has the topology of a disk 
times a line.
The boundary of spacetime at $\rho=\pi/2$ is a cylinder $S^1 \times R$.
(See Fig.~\ref{fig:one}) The bulk-boundary correspondence
asserts that a conformal field theory on this cylinder is dual to quantum
gravity in the bulk.

\begin{figure}                                 
\begin{center}                                 
\leavevmode                                 
\epsfxsize=5.5in
\epsfysize=3.5in                                 
\epsffile{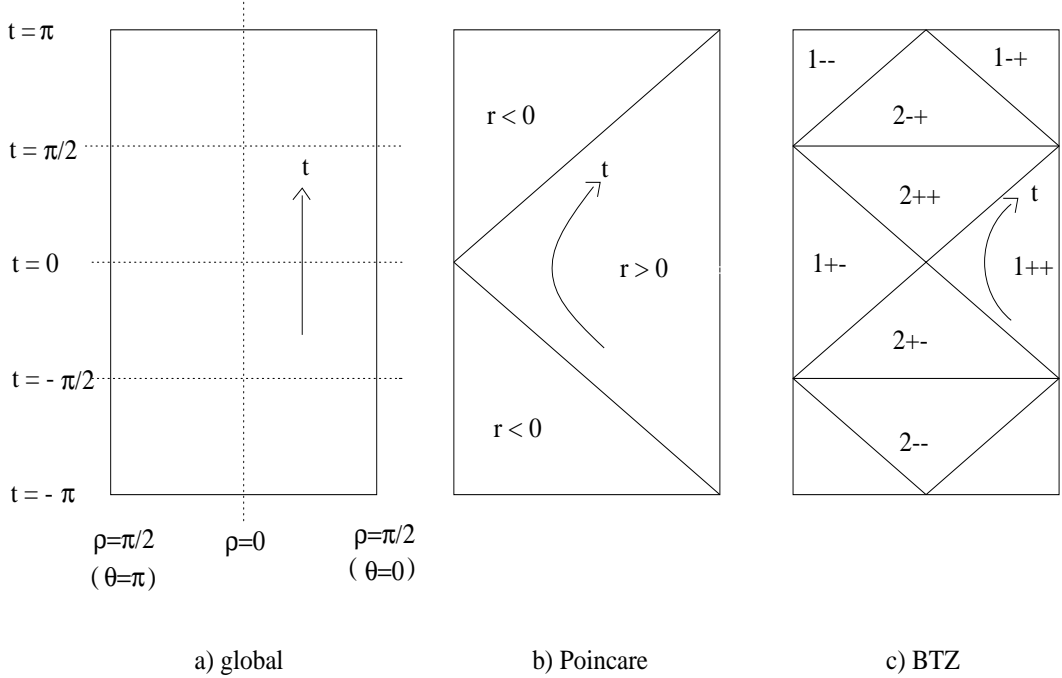}                                 
\end{center}                                 
\caption{Penrose diagrams for anti-de Sitter space.  Displayed are
vertical cross sections which cut through the center of the AdS cylinder.
In each figure, regions demarcated by solid lines identify the portion of
the spacetime covered by a single coordinate patch.  a) Global coordinates.
The boundary of the region is the surface of a cylinder.  b) Poincare
coordinates.  Here AdS is divided into two patches, with the two boundaries
at $r=0$ being conformal to flat Minkowski space. At the horizons, $r=\pm
\infty$.  c) BTZ coordinates.  AdS is divided into twelve patches, eight of
which appear in the two dimensional slice shown.  The eight boundaries are
each conformal to flat Minkowski space.\label{fig:two}}
\end{figure}

A Penrose diagram illustrating the causal structure can be drawn by drawn
by considering a two dimensional cross section of the hyperboloid.  We
choose to display a $X=0$ slice, and obtain the diagram in
Fig. ~\ref{fig:two}a.

\paragraph{Poincar\'e coordinates: }   Poincar\'e coordinates are defined by:
\begin{eqnarray}
U &=& {1 \over 2 r} (\cosm^2 + x^2 + r^2 - t^2) 
~~~~~~~~~;~~~~~~~
V =  \cosm\frac{t}{r} \nonumber \\
Y &=& {-1\over2r} (-\cosm^2 + x^2 + r^2 - t^2)
~~~~~~;~~~~~~ 
X =  \cosm\frac{x}{r} 
\label{poincare}
\end{eqnarray}
giving the metric:
\begin{equation}
ds^2 = (\cosm^2/r^2) ( -dt^2 + dx^2 + dr^2 )
\end{equation}
Here $t$ and $x$ range between $-\infty$ and $\infty$, and
$0\leq r \leq \infty$.  Poincar\'e coordinates only cover one half of
$\ads{3}$, as shown in Fig.~\ref{fig:one}.  There is a
horizon in these coordinates at $r =\infty$.  The boundary at $r= 0$
is clearly conformal to flat Minkowski space $R^{1,1}$.  The bulk-boundary
correspondence asserts that a conformal field theory on this boundary plane
is dual to quantum gravity in the bulk.  Note however, that the plane only
covers half of the cylindrical boundary of global $\ads{3}$.

The second half of $\ads{3}$ displayed in Fig.~\ref{fig:one} can be covered
by labelling the hyperboloid as in (\ref{poincare}) but now letting
$-\infty \le r \le 0$.  These two patches together cover $\ads{3}$, while
to cover ${\rm CAdS}_3$ one assembles a vertical tower of such patches.
The Penrose diagram is displayed in Fig. ~\ref{fig:two}b.

\paragraph{BTZ coordinates:}

For completeness, we consider a third coordinate system which is useful for
constructing the BTZ black hole.  Divide the hyperboloid into three regions:  
\begin{eqnarray}
{\rm Region \ }1:&\mbox{}&-U^2+X^2\le 0,\quad\quad -V^2+Y^2\ge 0 \nonumber \\
{\rm Region \ }2:&\mbox{}&-U^2+X^2\le 0 ,\quad\quad -V^2+Y^2\le 0 \nonumber \\
{\rm Region \ }3:&\mbox{}& -U^2+X^2\ge 0,\quad\quad -V^2+Y^2\le 0  \
\label{boosted coordinates}
\end{eqnarray}
Note that one cannot take both $-U^2+X^2$ and $-V^2+Y^2$ to be positive.
We then cover each region with four coordinate patches. 
\begin{eqnarray}
U&=& \pm \hat{r} \cosh{\hat{\phi}} ~~~~~~;~~~~~~
V= \sqrt{{\hat{r}}^2-\cosm^2}  \sinh{\hat{t}}\nonumber \\
X&=& \hat{r} \sinh {\hat{\phi}} ~~~~~~~~~;~~~~~~ 
Y= \pm  \sqrt{{\hat{r}}^2-\cosm^2}  \cosh{\hat{t}} ~~~~~~~~~~~~{\rm Region \ }1
\nonumber \\
ds^2&=& -({\hat{r}}^2-\cosm^2)d{\hat{t}}^2 +\cosm^2({\hat{r}}^2-\cosm^2)^{-1}d{\hat{r}}^2
+{\hat{r}}^2 d{\hat{\phi}}^2 \
\label{region 1}
\end{eqnarray}  

\begin{eqnarray}
U&=& \pm \hat{r} \cosh{\hat{\phi}} ~~~~~~;~~~~~~
V= \pm\sqrt{\cosm^2-{\hat{r}}^2}  \cosh{\hat{t}}\nonumber \\
X&=& \hat{r} \sinh {\hat{\phi}} ~~~~~~~~~;~~~~~~
Y= \sqrt{\cosm^2-{\hat{r}}^2}  \sinh{\hat{t}} ~~~~~~~~~~~~{\rm Region \ }2
\nonumber \\
ds^2&=& -({\hat{r}}^2-\cosm^2)d{\hat{t}}^2 +\cosm^2({\hat{r}}^2-\cosm^2)^{-1}d{\hat{r}}^2
+{\hat{r}}^2 d{\hat{\phi}}^2 \
\label{region 2}
\end{eqnarray} 

\begin{eqnarray}
U&=& \hat{r} \sinh{\hat{\phi}} ~~~~~~~~~;~~~~~~
V= \pm\sqrt{{\hat{r}}^2+\cosm^2}  \cosh{\hat{t}}\nonumber \\
X&=& \pm\hat{r} \cosh{\hat{\phi}} ~~~~~~;~~~~~~
Y= \sqrt{{\hat{r}}^2+\cosm^2}  \sinh{\hat{t}} ~~~~~~~~~~~~{\rm Region \ }3
\nonumber \\
ds^2&=& ({\hat{r}}^2+\cosm^2)d{\hat{t}}^2 +\cosm^2({\hat{r}}^2+\cosm^2)^{-1}d{\hat{r}}^2
-{\hat{r}}^2 d{\hat{\phi}}^2 \
\label{region 3}
\end{eqnarray}

In all three regions, $\hat{t}$ and $ \hat{\phi}$ range between $-\infty$ and
$\infty$. $\hat{r}$ has range $\cosm\le\hat{r} \le \infty$ in region 1, 
$0\le \hat{r} \le \cosm$ in region 2, and $0\le\hat{r} \le \infty$ in region 3.
So altogether there are twelve patches
covering $\ads{3}$.  To draw a Penrose diagram we again consider the slice
$X=0$.  Note  that on this slice region 3 is a dimension one submanifold,
while regions 1 and 2 are dimension two submanifolds.  Thus only regions 1 and
2 will appear in the Penrose diagram, and we label the various patches as
$1\pm\pm$, $2\pm\pm$ in an obvious notation.  The Penrose diagram then appears
as in Fig.~\ref{fig:two}c.  

To make a BTZ black hole \cite{BTZ} of mass $M$ from the above coordinates
one simply makes $\hat{\phi}$ periodic with period $2\pi \sqrt{M}$.

\subsection{$\ads{d+1}$}
The various coordinate systems defined above generalize straightforwardly to 
arbitrary dimensions.  Global coordinates for $\ads{d+1}$ give the metric:
\begin{equation}
ds^2 = \cosm^2 [-\sec^2\rho \, dt^2 + \sec^2\rho \, d\rho^2 + \tan^2\rho \,
d\Omega_{d-1}^2] 
\end{equation}
with $0\leq \rho \leq \pi/2$, $-\infty\leq t\leq \infty$.  Thus $\ads{d+1}$
is globally a d-dimensional disk times a line and the boundary at $\rho =
\pi/2$ is a cylinder $S^{d-1} \times R$.    The d'Alembertian operator in global
coordinates is
\begin{equation}
\cosm^2 \Box=-\cos^2{\rho}\partial_t^2+\cos^2{\rho}\partial_\rho^2 +(d-1)\cot{\rho}
\partial_\rho+\cot^2{\rho}\del^2_{ S^{d-1}}.
\end{equation}

Poincar\'e coordinates yield a metric:
\begin{equation}
ds^2 = (\cosm^2/r^2) (-dt^2  + d\vec{x}^2 + d r^2 )
\end{equation}
Here $d\vec{x}^2$ is the flat metric on $R^{d-1}$ and $0\leq r\leq\infty$.
There is a horizon at $r = \infty$ and the boundary at $r=0$ is the plane
$R^{d-1,1}$.  The d'Alembertian operator in Poincar\'e coordinates is
\begin{equation}
\cosm^2 \Box= -r^2\partial_t^2 + r^2\partial_r^2-(d-1)r\partial_r +r^2\del_{R^{d-1}}^2.
\end{equation}


\end{document}